\documentclass[aps,pra,twocolumn,showpacs,superscriptaddress,floatfix]{revtex4}
\usepackage{graphicx}
\usepackage{bm}
\begin{document}
\title{Manipulating ultracold polar molecules with microwave radiation: \\
the influence of hyperfine structure}
\author{J. Aldegunde}
\email{E-mail: Jesus.Aldegunde@durham.ac.uk}%
\affiliation{Department of Chemistry, Durham University, South Road,
Durham, DH1~3LE, United Kingdom}
\author{Hong Ran}
\affiliation{Department of Chemistry, Durham University, South Road,
Durham, DH1~3LE, United Kingdom}
\affiliation{Institute of
Theoretical and Computational Chemistry, Key Laboratory of
Mesoscopic Chemistry, School of Chemistry and Chemical Engineering,
Nanjing University, Nanjing 210093, China}
\author{Jeremy M. Hutson}
\email{E-mail: J.M.Hutson@durham.ac.uk}%
\affiliation{Department of Chemistry, Durham University, South Road,
Durham, DH1~3LE, United Kingdom}

\date{\today}

\begin{abstract}
We calculate the microwave spectra of ultracold KRb alkali
metal dimers, including hyperfine interactions and in the
presence of electric and magnetic fields. We show that
microwave transitions may be used to transfer molecules between
different hyperfine states, but only because of the presence of
nuclear quadrupole interactions. Hyperfine splittings may also
complicate the use of ultracold molecules for quantum
computing. The spectrum of molecules oriented in electric
fields may be simplified dramatically by applying a
simultaneous magnetic field.
\end{abstract}


\maketitle

Ultracold molecules offer striking new possibilities in many
areas of science \cite{Krems:book:2009}. The novel applications
include the development of quantum control schemes using
electric and magnetic fields, the employment of ultracold
molecules in quantum information storage and processing, the
production of strongly interacting quantum gases and the
possibility of performing precision measurements of physical
quantities \cite{Carr:2009}.

It has been possible since 2003 to produce alkali metal dimers
in highly excited vibrational states in ultracold atomic gases
\cite{Hutson:IRPC:2006}. However, it is only in the last year
that it has been possible to transfer these molecules
coherently to deeply bound states. This has now been achieved
for KRb \cite{Ni:KRb:2008}, Cs$_{2}$ \cite{Danzl:v73:2008,
Mark:2009, Danzl:ground:2009} and triplet Rb$_{2}$
\cite{Lang:ground:2008}. RbCs \cite{Sage:2005}, LiCs
\cite{Deiglmayr:2008} and NaCs \cite{Haimberger:2009} have also
been prepared in deeply bound states, but so far by incoherent
methods.

The bound states of diatomic molecules are described by
vibrational and rotational quantum numbers $v$ and $N$.
However, this does not suffice to specify the state completely,
as most molecules also possess complicated hyperfine structure
\cite{Townes:1975, Brown:2003}, even for singlet molecules in
$N$=0 states \cite{Aldegunde:2008, Aldegunde:2009}. This
structure cannot be neglected in ultracold studies, both
because hyperfine energy splittings can be of the same order of
magnitude as the thermal energy and because the molecules must
be in the same hyperfine state to achieve Bose-Einstein
condensation or Fermi degeneracy.

The interaction of cold molecules with microwave fields plays a
central role in proposals for microwave traps
\cite{DeMille:trap:2004, Alyabyshev:2009}, for tuning
molecule-molecule interactions to form novel quantum phases
\cite{Buechler:2007, Micheli:2007, Gorshkov:2008}, for studying
the dynamics of quantum phase transitions \cite{Micheli:2006,
Wall:2009} and for the employment of ultracold molecules in
quantum computing \cite{DeMille:2002, Andre:2006,
Micheli:2006}. The goal of this work is to show that the
hyperfine structure of microwave molecular spectra is important
in experiments involving cold and ultracold molecules.

In the present work, we simulate the microwave spectrum of an
alkali metal dimer including hyperfine interactions and in the
presence of electric and magnetic fields. We consider
$^{40}{\rm K}^{87}{\rm Rb}$ in the ground electronic state
($^{1}\Sigma ^{+}$), which has been experimentally prepared in
the ground rovibrational state ($v$=0, $N$=0) by Ni {\em et
al.} \cite{Ni:KRb:2008}. However, the spectra of other alkali
metal dimers will display very similar features. We demonstrate
the importance of the hyperfine structure by discussing (i) how
microwave transitions can be used to transfer molecules between
hyperfine states and (ii) the consequences of hyperfine
structure for the use of ultracold polar molecules in quantum
computing according to the scheme proposed by DeMille
\cite{DeMille:2002}.

The first step in the experiment of Ni {\em et al.}\
\cite{Ni:KRb:2008} is to form $^{40}{\rm K}^{87}{\rm Rb}$
dimers in a high-lying vibrational state from the corresponding
ultracold atoms. This is accomplished by tuning the magnetic
field across a Feshbach resonance. The resulting dimers are
characterized by a projection of the total angular momentum on
the direction of the field $M_{\rm F}$=$-7/2$. They are then
transferred into the ground rovibrational state using STIRAP
(STImulated Raman Adiabatic Passage) \cite{Bergmann:1998}. The
transfer is carried out at a magnetic field $B$=545.9 G. The
polarization of the lasers is such that $M_{\rm F}$ is
conserved during the STIRAP transfer. We therefore concentrate
here on the $M_{\rm F}$=$-7/2$ levels for $v$=0, $N$=0 and 1,
although by selecting a different initial state and/or by
changing the polarization of the STIRAP lasers it would be
possible in principle to populate different hyperfine states.

The molecular Hamiltonian of a $^{1}\Sigma$ molecule in the
presence of external fields can be written \cite{Brown:2003,
Aldegunde:2008, Aldegunde:2009}
\begin{equation}\label{eqH}
H=H_{\rm{rot}}+H_{\rm{hf}}+H_{\rm{S}}+H_{\rm{Z}}
\end{equation}
where $H_{\rm{rot}}$, $H_{\rm{hf}}$, $H_{\rm{S}}$ and
$H_{\rm{Z}}$ represent the rotational, hyperfine, Stark and
Zeeman contributions respectively. We construct and diagonalize
the Hamiltonian matrix in an uncoupled basis set $|N,
M_{\rm{N}}\rangle  |I_{\rm{K}} M_{\rm{K}}\rangle |I_{\rm{Rb}}
M_{\rm{Rb}} \rangle$, where $N$ is the molecular rotational
angular momentum and $I_{\rm{K}}$=4 and $I_{\rm{Rb}}$=3/2 are
the nuclear spins. $M_{\rm{N}}$, $M_{\rm{K}}$ and $M_{\rm{Rb}}$
are the projection quantum numbers for $N$, $I_{\rm{K}}$ and
$I_{\rm{Rb}}$ on the axis defined by the magnetic field. If the
electric and magnetic fields are parallel, the total projection
$M_{\rm F}$=$M_{\rm{N}}$+$M_{\rm{K}}$+$M_{\rm{Rb}}$ is a good
quantum number. Since $I_{\rm K}$ and $I_{\rm Rb}$ are fixed,
we abbreviate the basis functions to $|N, M_{\rm{N}},
M_{\rm{K}}, M_{\rm{Rb}} \rangle$. The matrix elements of the
various terms in the Hamiltonian and of the transition dipole
operator in this basis set are obtained by standard angular
momentum techniques \cite{Zare, Aldegunde:2008}.

The hyperfine and Zeeman Hamiltonians consist of several terms
whose coupling constants have been evaluated using DFT
calculations with relativistic corrections: see reference
\cite{Aldegunde:2008} for details of the methods used for the
calculation of the coupling constants. The Stark Hamiltonian is
evaluated using the experimental value of the KRb dipole
moment, $\mu$=0.566~D \cite{Ni:KRb:2008}.

\begin{figure}[t]
\includegraphics[width=85mm]{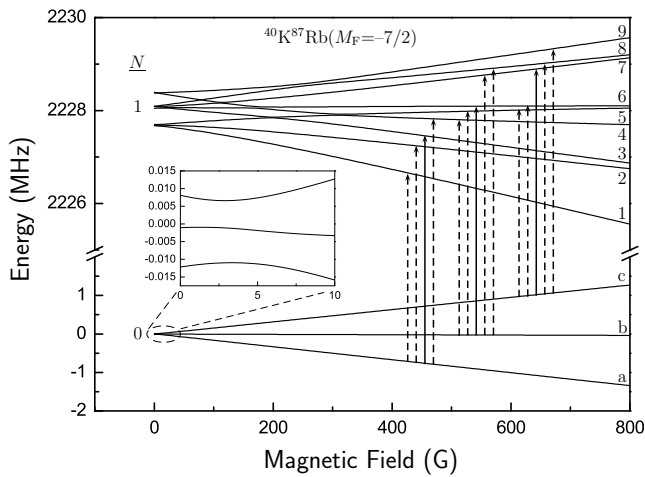}
\caption{Hyperfine and Zeeman energy levels for $M_{\rm F}$=$-7/2$
states of $^{40}$K$^{87}$Rb $(v$=0, $N=0-1$) in zero electric field.
The more probable microwave transitions from $N$=0 (a-c) to $N$=1
(1-9) states at a magnetic field $B$=545.9 G are
indicated with arrows (see Fig.\ \ref{figsp} for more details).
Although all the arrows correspond to transitions at $B$=545.9 G,
they have been displaced in order to make visualization easier.
Continuous lines represent the most probable transition for each
$N$=0 state. The small panel contains a blow-up of the $N$=0 levels
for values of the magnetic field below 10 G.} \label{figel}
\end{figure}

The hyperfine and Zeeman splitting for the $M_{\rm F}$=$-7/2$
states of $^{40}$K$^{87}$Rb ($N$=0 and 1) in zero electric
field is shown in Fig.\ \ref{figel}. All the apparent crossings
between energy levels are avoided crossings. Three terms in the
Hamiltonian (apart from $H_{\rm{rot}}$) determine the main
features of this figure: the scalar nuclear spin-spin, nuclear
electric quadrupole, and nuclear Zeeman interactions
\cite{Aldegunde:2008, Aldegunde:2009}. For the $N$=0 levels,
the zero-field splitting arises from the scalar part of the
electron-mediated interaction between the magnetic moments of
the nuclei, characterized by coupling constant $c_4$=$-2.0304$
kHz. The zero-field splitting amounts to 20 kHz, as shown in
the small panel in Fig.\ \ref{figel}. For the $N$=1 levels, the
zero-field splitting is mainly due to the nuclear electric
quadrupole term, with coupling constants $(eQq)_{\rm
^{40}K}$=$0.306$ MHz and $(eQq)_{\rm ^{87}Rb}$=$1.520$ MHz, and
amounts to approximately 1 MHz. This term, which describes the
interaction between the nuclear quadrupole moments and the
electric field gradient created by the electrons at the nuclear
positions, is much larger than the scalar spin-spin interaction
for $^{40}$K$^{87}$Rb and most of the other alkali dimers. It
generally dominates the zero-field splitting except for $N$=0,
because the matrix elements of the nuclear electric quadrupole
hamiltonian between $N$=0 basis functions vanish. The main
contribution to the Zeeman Hamiltonian comes from the
interaction between the nuclear magnetic moments and the
magnetic field (nuclear Zeeman effect). Other terms, such as
the tensor nuclear spin-spin, nuclear spin-rotation and
rotational Zeeman interactions, are much less significant.

\begin{figure}[tb]
\includegraphics[width=70mm]{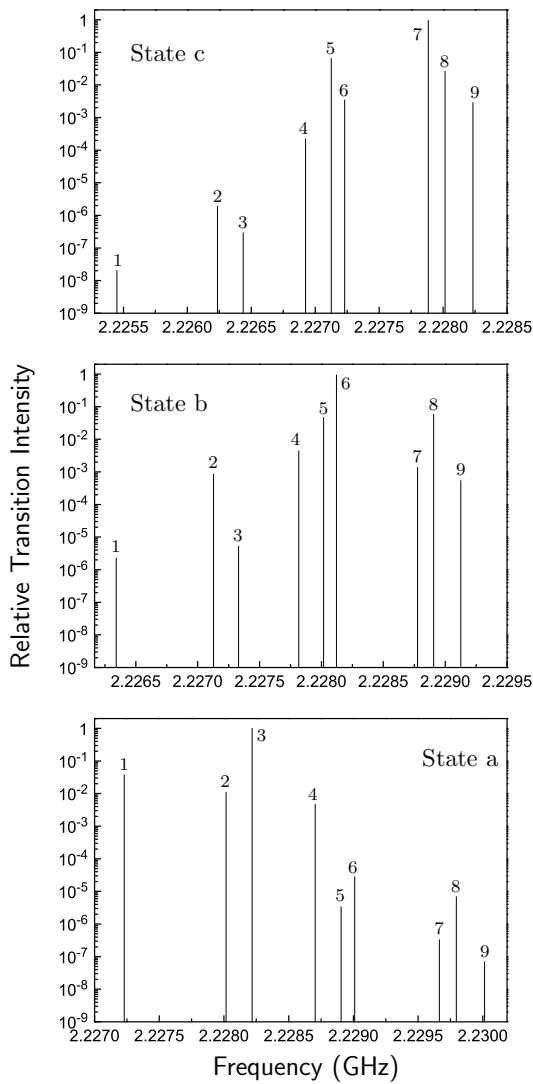}

\caption{Relative probability for the
$^{40}$K$^{87}$Rb$(v$=$0,M_{\rm F}$=$-7/2)$ microwave
transitions from $N$=0 to $N$=1 hyperfine states at $B$=545.9 G
and zero electric field. The transitions shown here are for
microwave radiation polarized parallel to the magnetic field,
so have $\Delta M_{\rm F}$=0. Each panel contains all the
transitions from one of the three $N$=0 hyperfine states,
labeled as a (bottom), b (middle) and c (top). The $N$=1 states
are labeled 1 to 9. The process with the largest transition
dipole moment (peak 3 in the bottom panel) has been assigned an
intensity of 1 and all the other transitions are relative to
it.} \label{figsp}
\end{figure}

It is important to develop methods to control the hyperfine
states of ultracold molecules. It may be possible to use
microwave transitions between $N=0$ and $N$=1 levels to
transfer alkali dimers between different hyperfine states, as
has been done for cold ND$_3$ and OH molecules
\cite{vanVeldhoven:2005, Hudson:2006}. In order to establish
whether this is feasible, we simulate the microwave spectrum
for all three $N$=0 hyperfine states of $^{40}$K$^{87}$Rb with
$M_{\rm F}$=$-7/2$. We evaluate the transition dipole moments
of the relevant transitions in the presence of a magnetic field
$B$=545.9 G. The relative intensities for a microwave field
polarized parallel to the magnetic field are shown in Fig.\
\ref{figsp}. Numerical values of the level energies and
intensities for both parallel and non-parallel polarization are
available as supplementary material
\cite{KRb-hyperfine-spectra-EPAPS}. For each of the $N$=0
hyperfine states, there are several transitions with intensity
within a factor of $10^3$ of that of the strongest transition,
highlighted with arrows in Fig.\ \ref{figel}. Some of the $N$=1
hyperfine states can be reached with significant intensity from
more than one $N$=0 state. Microwave radiation with a
polarization that is not parallel to the magnetic field can
also drive transitions with $\Delta M_{\rm F} \ne 0$. It will
therefore be possible to use microwave transitions to transfer
ultracold alkali metal dimers between hyperfine states,
including to the $M_{\rm F}$=$-5/2$ absolute ground state.

It is important to note that the transfer would {\em not} be
possible in the absence of nuclear quadrupole interactions. If
the quadrupole terms are omitted, the subsidiary transitions in
Fig.\ \ref{figsp} have intensities at least 6 orders of
magnitude less than that of the main peaks. The selection rules
for transitions driven by a $z$-polarized microwave field are
$\Delta N$=$\pm 1$, $\Delta M_{\rm{N}}$=0, $\Delta
M_{\rm{K}}$=0 and $\Delta M_{\rm{Rb}}$=0 in the uncoupled basis
set. At $B$=545.9 G, the $N$=0 block of the Hamiltonian is
dominated by the nuclear Zeeman term, which is diagonal in this
basis set. The only interactions that couple $N$=0 to higher
$N$ are small hyperfine terms. Because of this, $M_{\rm{N}}$,
$M_{\rm{K}}$ and $M_{\rm{Rb}}$ are nearly good quantum numbers
for $N$=0. If this were also the case for the $N$=1 levels,
only one transition from each initial state,
\begin{equation}
|N=0, M_{\rm{N}}=0, M_{\rm{K}}, M_{\rm{Rb}} \rangle \rightarrow
|N=1, M_{\rm{N}}=0, M_{\rm{K}}, M_{\rm{Rb}} \rangle,
\end{equation}
would have significant intensity in Fig.\ \ref{figsp}. However,
the electric quadrupole interaction prevents this: it is
off-diagonal in the uncoupled basis set for $N$=1 and strongly
mixes the $|N$=$1, M_{\rm{N}}, M_{\rm{K}}, M_{\rm{Rb}} \rangle$
basis functions. Since the projection quantum numbers are not
well defined for $N$=1, the selection rules involving them are
less restrictive.

\begin{figure*}[tb]
\includegraphics[width=180mm]{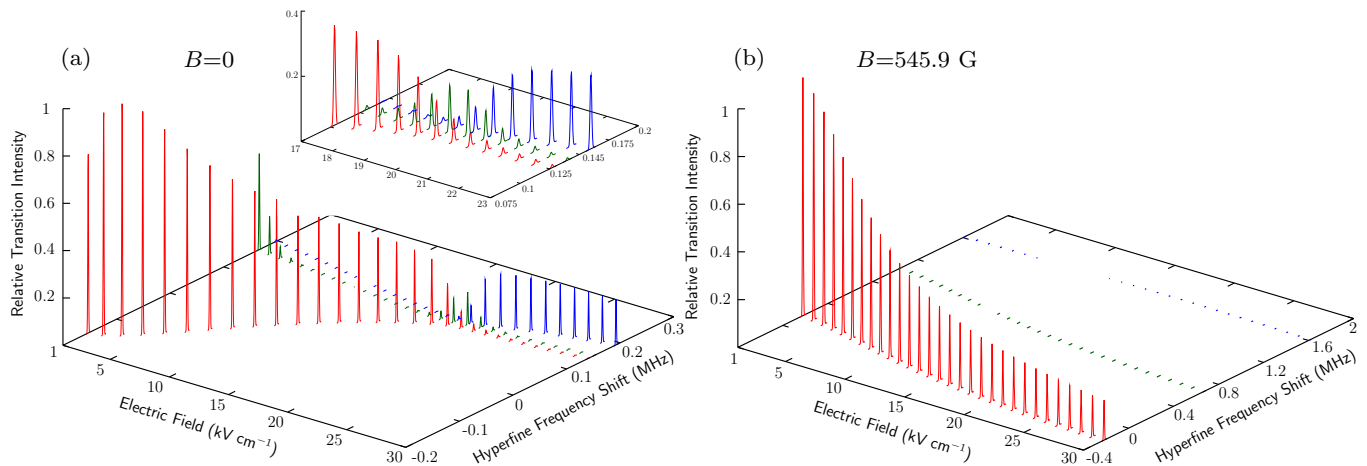}
\caption{(Color online) Relative intensities of the microwave
transitions from the lowest $v$=0,$N$=0 hyperfine state of
$^{40}$K$^{87}$Rb to the three possible $N$=1, $M_{\rm{N}}$=0
hyperfine states, as a function of the external electric field, (a)
in the absence of a magnetic field and (b) in the presence of a
parallel magnetic field $B$=545.9 G. All levels have $M_{\rm
F}$=$-7/2$. The microwave field is polarized parallel to the fields.
The inset shows an enlargement of the crossing region in (a). The
most intense transitions in (a) and (b) have been assigned a peak
intensity of 1. In order to keep the spectrum on a single frequency
scale as a function of electric field, we plot the hyperfine
frequency shift instead of the absolute frequency. This is obtained
by subtracting the frequency for the transition in the absence of
hyperfine splittings.} \label{figsp3d}
\end{figure*}

Hyperfine structure will also be important in applications of
ultracold molecules to quantum computing. DeMille
\cite{DeMille:2002} has proposed a design for a quantum
computer in which the qubits are formed from ultracold polar
molecules held in a 1D optical lattice. Each of the trap sites
($10^{4}$ in the original design) is occupied by a single
molecule. To facilitate individual addressing, an external
electric field $\epsilon$ that varies linearly with the
position in the array is applied. The $|0\rangle$ and
$|1\rangle$ states of the qubits are field-induced mixtures of
the $M_{\rm{N}}$=0 rotational states, characterized by
orientations of the molecular electric dipole parallel and
antiparallel to the electric field respectively. At the fields
where the device operates, $|0\rangle$ is predominantly $N$=0
and $|1\rangle$ is predominantly $N$=1. Switching between the
$|0\rangle$ and $|1\rangle$ states is driven by microwave
fields whose polarization is parallel to $\epsilon$ ($\Delta
M_{\rm{F}}$=0).

In DeMille's design, the qubits are KCs molecules. When the
original experimental parameters are adapted to use KRb, the
optimum range for the external electric field is from
approximately 7 to 18 kV/cm and the electric resonance
frequencies needed to address the molecules range from 3.5 to 6
GHz in steps of 250 kHz.

The ultracold dimers will properly represent qubits only if it
is possible to switch repeatedly between the $|0\rangle$ and
$|1\rangle$ states of one molecule without populating other
states and without modifying the state of any other molecule
during the process. The existence of several possible hyperfine
transitions can complicate the operation of the device by
making the individual addressing of molecules more difficult.

In order to determine the extent of these difficulties, we
simulate the microwave spectrum for transitions from the lowest
$N$=0 hyperfine state of KRb in an electric field. Even an
electric field as small as 0.5 kV/cm is sufficient to separate
the levels for $M_{\rm N}$=0 and $\pm1$. Fig.\ \ref{figsp3d}(a)
shows the hyperfine frequency shifts and intensities for
transitions to the three $N$=1,$M_{\rm{N}}$=0 levels for
$M_{\rm F}$=$-7/2$, calculated for electric fields between 1
and 30 kV/cm in the absence of a magnetic field. The peaks
display a crossing as a function of the electric field
$\epsilon$. The crossing region, corresponding to values of
$\epsilon$ between 17 and 23 kV/cm, is shown in the inset of
Fig.\ \ref{figsp3d}(a). It is characterized by the existence of
three significant peaks for each value of the field. The
behavior of the peaks reflects that of the $N$=1,$M_{\rm{N}}$=0
hyperfine levels, which display avoided crossings in the same
range of electric fields. Even outside the crossing region, all
three peaks have significant intensities (mostly $\geq10^{-4}$
of that of the main peak).

Hyperfine splittings will thus complicate the individual
addressing of molecules in a quantum computer based on
DeMille's design \cite{DeMille:2002}. For electric fields
between 7 and 15 kV/cm, the peaks spread over a range of
frequency shifts that is comparable to or larger than the
frequency step used for addressing (250 kHz). This may cause
overlapping between the spectra of neighboring molecules.

This difficulty can be overcome by applying a magnetic field
parallel to the electric field. The magnetic field resolves the
near-degeneracy between levels with different values of $M_{\rm
K}$ and $M_{\rm Rb}$ but the same $M_{\rm F}$. The spectrum
calculated for $B$=545.9 G (the magnetic field value in the
experiment of Ni {\em et al.}\ \cite{Ni:KRb:2008}) is shown in
Fig.\ \ref{figsp3d}(b). A single transition dominates the
spectrum for all values of the electric field and no crossing
is displayed. The subsidiary transitions are 7 or 8 orders of
magnitude weaker than the primary transition.

It is thus clear that a detailed understanding of hyperfine
structure is essential when designing experiments that involve
microwave transitions in ultracold molecules. We have shown
that microwave transitions could be used to transfer polar
molecules between hyperfine states, but only because of the
presence of the nuclear electric quadrupole interaction. We
have also investigated the possibility of using ultracold polar
molecules as the basis for the logic gates of a quantum
computer. In this case the hyperfine splittings introduce some
difficulties in the operation of the device, but these can be
overcome by applying a magnetic field as well as an electric
field.

We are grateful to Jun Ye for discussions about the current KRb
experiments and to EPSRC for funding of the collaborative project
QuDipMol under the ESF EUROCORES programme EuroQUAM. HR is grateful
to the China Scholarship Council for funding her joint PhD student
program in Durham.

{\em Note added.} --- After this calculations described here
were carried out, Ospelkaus {\em et al.} \cite{Ospelkaus:2009}
succeeded in using microwave transitions to transfer molecules
between the $N$=0 hyperfine states of $^{40}$K$^{87}$Rb.

%\bibliography{../../all,./jesus}

\end{document}